\documentclass[a4paper,12pt]{article}
\pdfoutput=1
\usepackage{feynmp-auto,expdlist}
\usepackage{amsmath, amsfonts, amssymb}
\usepackage{graphicx}
\usepackage{enumerate}
\usepackage{hyperref}
\usepackage{latexsym}
\usepackage{dsfont}
\usepackage{hepnicenames}
\usepackage{enumerate}
\usepackage{soul}
\usepackage[normalem]{ulem}
\usepackage{bbold}

\usepackage{units}

\newcommand{\es}[2] {\begin{equation} \label{#1} \begin{split} #2 \end{split} \end{equation}}

\usepackage{mathrsfs,graphicx,rotating,amsmath,amsfonts,mathtools,booktabs,amssymb,wasysym}
\usepackage{hyperref}\usepackage{slashed}
\usepackage[nosort]{cite}
\usepackage[table,xcdraw,dvipsnames]{xcolor}
\usepackage{bm}
\usepackage{graphicx}
\usepackage{multirow,multicol}
\hypersetup{colorlinks,bookmarksopen,bookmarksnumbered,
	linkcolor=blus,pdfstartview=FitH,urlcolor=rossos,citecolor=verde}
\allowdisplaybreaks

\usepackage{mathrsfs}
\usepackage{tabularx}
\usepackage{amsmath}

\allowdisplaybreaks
\usepackage{multicol}
\usepackage{color}
\definecolor{rosso}{cmyk}{0,1,1,0.4}
\definecolor{rossos}{cmyk}{0,1,1,0.55}
\definecolor{rossoc}{cmyk}{0,1,1,0.2}
\definecolor{blu}{cmyk}{1,1,0,0.3}
\definecolor{blus}{cmyk}{1,1,0,0.6}
\definecolor{bluc}{cmyk}{1,1,0,0.1}
\definecolor{verde}{cmyk}{0.92,0,0.59,0.25}
\definecolor{verdec}{cmyk}{0.92,0,0.59,0.15}
\definecolor{verdes}{cmyk}{0.92,0,0.59,0.4}

\font\tenrsfs=rsfs10 at 12pt
\font\sevenrsfs=rsfs7
\font\fiversfs=rsfs5
\newfam\rsfsfam
\textfont\rsfsfam=\tenrsfs
\scriptfont\rsfsfam=\sevenrsfs
\scriptscriptfont\rsfsfam=\fiversfs

\def\circa#1{\,\raise.3ex\hbox{$#1$\kern-.75em\lower1ex\hbox{$\sim$}}\,}
\makeatletter

\font\ital=cmu10

\def\hhref#1{\href{http://arxiv.org/abs/#1}{arXiv:#1}}
\usepackage{xstring}
\newcommand{\hhrefq}[1]{\IfSubStr{#1}{:}{\href{http://inspirehep.net/search?ln=en&ln=en&p=#1&of=hb&action_search=Search&sf=&so=d&rm=&rg=25&sc=0}{InSpire:#1}}{\hhref{#1}}}

\def\art{\@ifnextchar[{\eart}{\oart}}
\def\eart[#1]#2#3#4#5#6{{\rm #2}, {\em #3 \bf #4} {\rm (#6) #5} ({\em #1})}
\def\article{\@ifnextchar[{\earticle}{\oarticle}}
\def\oarticle#1#2#3#4#5#6{{\rm #1}, {\ital ``#6''}, {\rm #2 #3 (#5) #4}}
\def\earticle[#1]#2#3#4#5#6#7{{\rm #2}, {\ital ``#7''}, {\rm #3 #4 (#6) #5}  [\hhrefq{#1}]}
\def\hepart[#1]#2{{\rm #2, \sl#1}}
\def\heparticle[#1]#2#3{#2, {\ital ``#3''} [\hhrefq{#1}]}
\newcommand{\doi}[1]{\href{http://dx.doi.org/#1}{[link]}}

\newcommand{\hhrefqq}[1]{\IfBeginWith{#1}{10.}{\href{https://doi.org/#1}{doi:#1}}{\hhrefq{#1}}}
\def\earticle[#1]#2#3#4#5#6#7{{\rm #2}, {\ital ``#7''}, {\rm #3 #4 (#6) #5}  [\hhrefqq{#1}]}

\newcounter{alphaequation}[equation]
\def\thealphaequation{\theequation\hbox to
	0.6em{\hfil\alph{alphaequation}\hfil}}
\def\eqnsystem#1{
	\def\@eqnnum{{\rm (\thealphaequation)}}
	\def\@@eqncr{\let\@tempa\relax \ifcase\@eqcnt \def\@tempa{& & &} \or
		\def\@tempa{& &}\or \def\@tempa{&}\fi\@tempa
		\if@eqnsw\@eqnnum\refstepcounter{alphaequation}\fi
		\global\@eqnswtrue\global\@eqcnt=0\cr}
	\refstepcounter{equation} \let\@currentlabel\theequation \def\@tempb{#1}
	\ifx\@tempb\empty\else\label{#1}\fi
	\refstepcounter{alphaequation}
	\let\@currentlabel\thealphaequation
	\global\@eqnswtrue\global\@eqcnt=0 \tabskip\@centering\let\\=\@eqncr
	$$\halign to \displaywidth\bgroup \@eqnsel\hskip\@centering
	$\displaystyle\tabskip\z@{##}$&\global\@eqcnt\@ne
	\hskip2\arraycolsep\hfil${##}$\hfil& \global\@eqcnt\tw@\hskip2\arraycolsep
	$\displaystyle\tabskip\z@{##}$\hfil
	\tabskip\@centering&\llap{##}\tabskip\z@\cr}
\def\endeqnsystem{\@@eqncr\egroup$$\global\@ignoretrue} \makeatother

\newcommand\be{\begin{equation}}
	\newcommand\ee{\end{equation}}
\newcommand\ba{\begin{eqnarray}}
	\newcommand\ea{\end{eqnarray}}\newcommand\eq{\begin{equation}}
	\newcommand\en{\end{equation}}

\definecolor{Gray}{gray}{0.95}

\def\bal#1\eal{\begin{align}#1\end{align}}

\begin{document}
\vspace{1.5cm}

\begin{center}
	{\Large \bf \color{rossos} Explore the Axion Dark Matter through the Radio \\[4mm] Signals from Magnetic White Dwarf Stars}\\[1cm]

	{\bf Jin-Wei Wang$^{a,b,c,\P}$	, Xiao-Jun Bi$^{d,e,\dagger}$, Run-Min Yao$^{d,e,\ddagger}$, Peng-Fei Yin$^{d,\S}$}\\[7mm]

	{\it $^a$ Scuola Internazionale Superiore di Studi Avanzati (SISSA), via Bonomea 265, 34136 Trieste, Italy}\\[1mm]
	{\it $^b$ INFN, Sezione di Trieste, via Valerio 2, 34127 Trieste, Italy}\\[1mm]
	{\it $^c$ Institute for Fundamental Physics of the Universe (IFPU), via Beirut 2, 34151 Trieste, Italy}\\[1mm]
	{\it $^d$ Key Laboratory of Particle Astrophysics, Institute of High Energy Physics,
		Chinese Academy of Sciences, Beijing, China}\\[1mm]
	{\it $^e$ School of Physical Sciences, University of Chinese Academy of Sciences, Beijing, China}\\[1mm]
	\vspace{0.5cm}
	{\large\bf\color{blus}}

\begin{quote}
\large
Axion as one of the promising dark matter candidates can be detected through narrow radio lines emitted from the magnetic white dwarf stars. Due to the existence of the strong magnetic field, the axion may resonantly convert into the radio photon (Primakoff effect) when it passes through a narrow region in the corona of the magnetic white dwarf, where the plasma frequency is equal to the axion mass. We show that for the magnetic white dwarf WD 2010+310, the future experiment SKA phase 1 with 100 hours of observation
can effectively probe the parameter space of the axion-photon coupling $g_{a\gamma}$ up to $\sim 10^{-12}~ \text{GeV}^{-1}$ for the axion mass range of $0.2 \sim 3.7~ \mu$eV. Note that in the low mass region ($m_a \lesssim 1.5 ~\mu\text{eV}$), the WD 2010+310 could give greater sensitivity than the neutron star RX J0806.4-4123.
\end{quote}

\vspace{1.2cm}
	
\thispagestyle{empty}
\bigskip
\end{center}

\noindent\rule{2cm}{0.9pt}
\begin{flushleft}
$\P$ jinwei.wang@sissa.it\\
$\dagger$ bixj@ihep.ac.cn\\
$\ddagger$ yaorunmin@ihep.ac.cn\\
$\S$ yinpf@ihep.ac.cn
\end{flushleft}
\newpage

\setcounter{footnote}{0}

\tableofcontents

\section{Introduction}\label{intro}
The existence of dark matter (DM) has been established by solid astrophysical and cosmological observations\cite{0608407,1807.06209}. For quite a long time the weakly interacting massive particles (WIMPs)
are regarded as the most promising DM candidates, because they can naturally explain the DM relic
density \cite{0404175,1003.0904,1703.07364}. However, so far no convincing dark matter signal has been found in the direct detection, indirect detection, and collider detection experiments. Furthermore, the limitations on
the couplings between the DM particles and standard model particles are becoming more and more stringent \cite{1003.0904,1709.00688,2004.04547}. In this case, the experimental searches for other DM candidates have thus attracted increasingly attention in recent years \cite{1811.07873,1907.10610}.

Among many other alternatives, the QCD axion, a light neutral pseudoscalar particle associated with the U(1) Peccei-Quinn symmetry \cite{Peccei:1977ur}, is one of the best options due to several excellent theoretical characteristics: (1) it can resolve the strong CP problem very well \cite{PhysRevLett.38.1440,PhysRevLett.40.223,PhysRevLett.40.279}; (2) it can explain the observed DM abundance \cite{PRESKILL1983127,ABBOTT1983133,DINE1983137}. For more details we refer the reader to the excellent reviews of axion physics \cite{1602.00039,2003.01100,1510.07633}.

Based on the possible couplings between axion and the electromagnetic sector, a number of experiments have been set up to search for axion DM signals.
These interactions predict two different phenomena: (1) the conversion between an axion particle and a photon under magnetic fields (so-called Primakoff effect \cite{Pirmakoff:1951pj}), e.g. axion helioscope \cite{1705.02290,1401.3233}, "light shining through a wall" experiments \cite{1004.1313,1302.5647}, and so on; (2) the photon birefringence under axion background \cite{1802.07273,1805.11753,1809.01656,1811.03525,2008.13662,2003.11015}. In this paper we focus on the former phenomenon.

The compact stars, e.g. magnetic white dwarf stars (MWDs) and neutron stars (NSs), are very promising probes to search for the axion DM, since these stars host strong magnetic fields, in which the axion can be converted into detectable photon signals. For example, there are studies in the literature using the X-ray observations of MWDs to detect the star-born axions \cite{1903.05088} and detecting the radio signals from axion DM conversion in the magnetospheres of NSs \cite{0711.1264,1804.03145,1803.08230,2008.01877,2011.05378,2004.06486}. However, the researches on the conversion of axion DM in magnetospheres of the MWDs have not been studied so far. Although the magnetic fields of the MWDs are usually weaker than NSs, they have the larger geometrical sizes. In addition, there are several MWDs near the earth with the distances smaller than 50 pc which also gives them an advantage for detection.

In this work we focus on the signals of axion DM from MWDs whose magnetic fields are at order of $10^7\sim 10^8$ G. With such a strong magnetic field, the axion DM may be converted into photons within the coronae of these MWDs.
In particular, when the axion mass $m_a$ is equal to the plasma frequency $\omega_p$, conversion probability can be enhanced greatly (called resonant conversion).
With the number density of plasma at the base of the MWD corona $ n_e \sim 10^{10}~\text{cm}^{-3}$, we have $\omega_p = \sqrt{4\pi \alpha_\text{em} n_e/m_e} \sim \mu\text{eV}$, which corresponds to a frequency of $\sim$ GHz.
Interestingly, this frequency happens to be in the sensitive region of the terrestrial radio telescopes, such as the Square Kilometer Array (SKA) that covers the $50 \sim 13800 ~\text{MHz}$ frequency band \cite{ska}. Therefore, we propose to use the radio telescopes to search for the axion DM in this mass range. This proposal can be regarded as a good supplement to the other axion detection experiment, such as ADMX \cite{PhysRevD.64.092003,PhysRevLett.104.041301,1405.3685} and CAST \cite{1705.02290,1307.1985,1503.00610}.

This paper is outlined as follows. In Sec. \ref{structure_WD} we introduce the distribution of plasma density in the coronae of the MWDs and their magnetic field structure. In Sec. \ref{conversion} we give a brief calculation of axion-photon conversion probability in the magnetic fields of MWDs. In Sec. \ref{radioflux} we calculate the radio flux density of some MWDs candidates as well as the constraints on the axion-photon coupling strength $g_{a\gamma}$ at the SKA. Conclusions and further discussions are given in Sec. \ref{conclusions}.

\section{The corona of the magnetic white dwarf and its magnetic field structure }\label{structure_WD}
The corona of MWD is suggested by several theories \cite{Zheleznyakov1984,Serber1990,Thomas1995ApJ}, but it has not been observed yet. The X-radiation searches can be used to set constraints on the parameters of the corona of MWD \cite{V.Zheleznyakov,Weisskopf_2007}. For example, the Chandra observation of the single cool MWD GD 356 sets limits on the plasma density of the hot corona as $n_{e0} < 4.4\times10^{11} ~\text{cm}^{-3}$ with the temperature of corona $T_\text{cor}\sim 10^7 ~\text{K}$ \cite{Weisskopf_2007}, while in Ref.\cite{V.Zheleznyakov} the upper limit on the plasma density is $n_{e0}\sim 10^{10} ~\text{cm}^{-3}$ with $T_\text{cor}\gtrsim10^6 ~\text{K}$ for the MWD G99-47 (WD 0553+053).  In the following sections, we show that the MWDs satisfying these constraints can be promising probes to detect the axion DM.


In this work, for the properties of the MWDs' coronae, we adopt the same assumptions as in Ref.\cite{V.Zheleznyakov}: (1) the corona is composed of
fully ionized hydrogen plasma uniformly covering the entire surface of the white dwarf; (2) the field-aligned temperature of the electrons $T_\text{cor}\sim 10^6~ \text{K}$ is a constant
throughout the corona. Under these conditions the distribution of the electron density at $r$ is described by the
barometric formula \cite{V.Zheleznyakov,Weisskopf_2007}
\ba
\label{ne}
n_e(r)=n_{e0} ~\text{exp}
\left(
-\frac{r-R_\text{WD}}{H_{\text{cor}}}
\right),
\ea
where $n_{e0}$ is the density at the base of the corona, $R_\text{WD}$ is the radius of the MWDs, and
\ba
\label{Hcor}
H_{\text{cor}}=\frac{2 k_\text{B} T_\text{cor}}{m_\text{p} g}=21.90 \left(\frac{T_\text{cor}}{10^6 ~\text{K}}\right) \left(\frac{M_\text{WD}}{M_\odot}\right)\left(\frac{R_\text{WD}}{10^4 ~\text{km}}\right)^{-2} \text{km}
\ea
is the scale height of the isothermal corona, $k_\text{B}$ is the Boltzmann constant, $m_\text{p}$ is the proton mass, $g$ is the free-fall acceleration at the surface of MWDs, and $M_\text{WD}$ is the mass of the MWDs.
By using the resonant conversion condition $m_a = \omega_p$, the resonant conversion radius $r_c$ can be solved as
\begin{align}
	\label{resr}
	r_{c} & = R_\text{WD} + 21.90 \times \left[2.634+\text{ln}\left(\frac{n_{e0}}{10^{10} ~\text{cm}^{-3}}\right) + \text{ln}\left(\frac{\mu \text{eV}^2}{m_a^2}\right)\right]\notag\\
	&\qquad\qquad\qquad~~~  \times\left(\frac{T_\text{cor}}{10^6 ~\text{K}}\right) \left(\frac{M_\text{WD}}{M_\odot}\right)\left(\frac{R_\text{WD}}{10^4 ~\text{km}}\right)^{-2} \text{km}.
\end{align}

Highly MWDs may give a very complex magnetic field structure \cite{Ferrario:2015oda}.
In this work, for simplicity we take the dipole configuration and assume that the WD rotation axis is parallel to the magnetization axis \cite{1903.05088}:
\begin{align}\label{eq:B_dipole}
	\bm B = \frac{B_0}{2} \, \frac{R_\text{WD}^3}{r^3} \, \Bigl( 3 (\hat{\bm m} \cdot \hat{\bm r}) \, \hat{\bm r} - \hat{\bm m} \Bigr)
	\qquad \text{for} \quad r > R_\text{WD}, \,
\end{align}
where $B_0$ is the value of the magnetic field at the MWD's surface in the direction of the magnetic pole,  $\bm{m} = 2\pi B_0 R_\text{WD}^3 \hat{\bm m}$ is the magnetic dipole moment, $\bm r = r \hat{\bm r}$ is the spatial coordinate, and $r = |\bm r|$ represents the distance from the center of the MWD. Clearly, we can see that the direction of $\bm B$ only depends on the $\theta$, which denotes the angle between $\hat{\bm m}$ and $\hat{\bm r}$, and its magnitude can be expressed as
\begin{align}\label{eq:B_dipole2}
	B= |\bm B| =\frac{B_0}{2} \, \frac{R_\text{WD}^3}{r^3} \sqrt{(3 ~\text{cos}\theta ~\text{sin}\theta)^2+(3 ~\text{cos}^2\theta-1)^2} \qquad \text{for} \quad r > R_\text{WD}. \,
\end{align}

\section{The axion-photon conversion probability in the magnetosphere of the magnetic white dwarf}\label{conversion}

Since the axion-photon conversion within the strong magnetic field of compact stats has been studied a lot in previous work, here we only give a general description as well as several necessary results but omit the most of intermediate steps. More detailed derivations can be found in Ref. \cite{PhysRevD.37.1237,1804.03145,1903.05088}.

Considering that the axion DM particle starts out non-relativistic ($\sim10^{-3} ~c$) far away from the MWDs and is accelerated as it moves toward the MWDs, we can approximate the axion's trajectory as radial. Due to the axion-photon coupling term $- \frac{1}{4} g_{a\gamma\gamma} \, a \, F_{\mu\nu} \tilde{F}^{\mu\nu}$ in the Lagrangian, the axion DM may be converted into the photon when it passes through the MWD's magnetosphere. More specifically, the axion on the way to the MWD could be converted into the photon, which  totally reflects back out because of the larger plasma frequency in the inner regions of MWD; the axion could also pass through the MWD and then is converted in the magnetosphere of the other side of the MWD.

Set up a coordinate system so that the radial direction of axion motion $\hat{\bm r}$ is the $z$ axis. Considering the dipole configuration of magnetic field, we can make sure that $\bm B$ is always in the $y$-$z$ plane by rotating the frame around the z axis. In the high-magnetization limit \cite{1804.03145}, the equation of motion of $\bm{E}$ and $a$ have some interesting features: (1) $E_x$ decouples from the equations; (2) $E_z$ is not dynamical and can also be removed from the system of equations; (3) $E_y$ is the only component that can mix with axion field.

\begin{figure}

	\centering
	$$\includegraphics[width=0.47\textwidth]{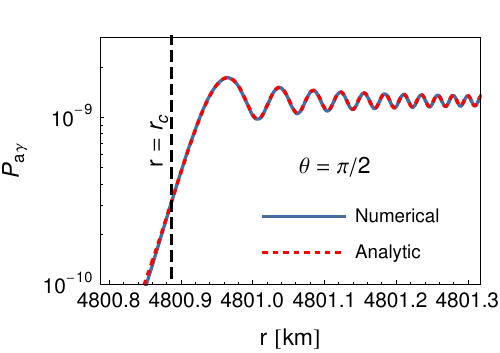}~~~~
	\includegraphics[width=0.47\textwidth]{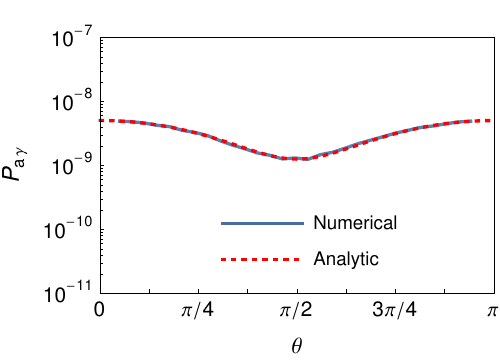}
	$$
	\vspace{-1cm}
	\caption{\em \label{fig:probabilityfigs}
		On the left: Energy transfer fraction $P_{a\gamma}$ as a function of distance $r$ with $\theta = \pi/2$, where the blue solid line is the numerical solution by solving eq. \eqref{ode}, red dashed line is the analytic approximation of eq. \eqref{finite_r1}, and black dashed line denotes the position of $r_c$. On the right: conversion probability $P_{a\gamma}$ as a function of $\theta$. Again, the analytical approximation of eq. \eqref{finite_r0} and numerical solution by solving eq. \eqref{ode} are shown here. Both results are from the MWD candidate WD 2010+310.}
		\label{figs1}
\end{figure}

Following \cite{1804.03145,PhysRevD.37.1237} we adopt the radial plane wave solution $a(r,t) = i e^{i \omega t - i k r} \tilde a(r)$ and $A_y(r,t) = e^{i \omega t - i k r} \tilde A_y(r)$, where $k = \sqrt{\omega^2 - m_a^2}$ represents the momentum of the axion. Under the temporal gauge $A_0=0$, we have $E_y = - dA_y/dt = -i\omega A_y$.
In fact, the resonant conversion happens in a very narrow region around $r_c$ at which the plasma frequency equals the axion mass (see Fig. \ref{fig:probabilityfigs}), so we can reasonably treat $k$ as a constant $k \simeq m_a v_c$\footnote{Here we have used the the non-relativistic approximation. For a typical MWD candidate in this work, the value of $v_c$ is $\sim 10^{-2} $.}, where $v_c$ is the axion velocity at $r_c$.
With the WKB approximations $|\tilde A_\parallel''(r)| \ll k |\tilde A'(r)|$ and $|\tilde a''(r)| \ll k |a'(r)|$, the mixing equations can be expressed as \cite{1804.03145}
\es{ode}{
	\left[ -i \frac{d}{dr} + \frac{1}{2 k}
	{
		\left(
		\begin{array}{cc}
			m_a^2 - \xi \, \omega_p^2   & -\Delta_B \\
			-\Delta_B & 0
		\end{array}
		\right)}
	\right]
	\left(\begin{array}{c}  \tilde A_y \\ \tilde a \end{array}\right)= 0 \,,
}
where
\es{}{
	\xi = {\sin^2 \tilde \theta \over 1 - {\omega_p^2 \over \omega^2} \cos^2 \tilde \theta } \,, \quad \, \Delta_B = B g_{a \gamma} \omega {\xi \over \sin \tilde \theta} \,,
}
$\tilde{\theta}$ is the angle between $\hat{\bm B}$ and $\hat{\bm r}$ (or $\hat{\bm z}$).
After diagonalizing the mixing matrix in eq. \eqref{ode}, we can easily prove that $|\tilde A_y(z)|^2 + |\tilde a(z)|^2 = \text{constant}$, which means that energy is conserved throughout the conversion process. We can define the energy transfer fraction as $p_{a \gamma}(r) = |\tilde{A}_y(r)|^2/|a_0|^2$ with the initial conditions $\tilde{A}_y(R_\text{WD}) = 0$ and $\tilde{a}(R_\text{WD}) = a_0$.

Following Ref. \cite{PhysRevD.37.1237}, we consider a perturbative solution of
eq. \eqref{ode}, which can be rewritten as a "Schrodinger equation" with the $r$ playing the role of time.
At the first order of $\Delta_B$, $p_{a \gamma}(r)$ can be derived from eq. \eqref{ode} as \cite{1804.03145,PhysRevD.37.1237}
\es{TProb_0}{
	p_{a \gamma}(r) = \frac{|\tilde{A}_y(r)|^2}{|a_0|^2} = \left| i \int_{R_\text{WD}}^r dr' \frac{\omega B(r')  g_{a \gamma \gamma} \xi(r')} {2 k \text{ sin}\tilde{\theta}}
	\times e^{{i f(r') \over 2 k}  } \right|^2,
}
where
\es{TProb_1}{
	f(r') =\int_{R_\text{WD}}^{r'} d\tilde r \big[m_a^2 - \xi(\tilde r)\omega_p^2(\tilde r) \big].
}
Note that the resonant conversion happens in a narrow region around $r_c$, so the choice of lower limit of the integral is irrelevant as long as the integral interval contains $r_c$. Since $1/2k \ll 1$, we can evaluate eq. \eqref{TProb_0} by the method of stationary phase. The probability of the axion-photon conversion at finite $r$ is given by
\es{TProb_finite_r}{
	p_{a  \gamma}(r)
	&\approx \frac{\xi(r_c)^2}{2v_c^2 \text{ sin}^2\tilde{\theta}} \, g_{a \gamma \gamma}^2 B(r_c)^2 L^2 \times G\left(\frac{r-r_c}{L}\right),
}
where
\begin{equation}
	L = \sqrt{\frac{2\pi m_a v_c}{|f''(r_c)|}}, \quad G(x) =\frac{\left(\frac{1}{2}+C(x)\right)^2 + \left(\frac{1}{2}+S(x)\right)^2}{2},
\end{equation}
which is defined by
in terms of the Fresnel $C$ and $S$ integrals.
Considering that all the MWD candidates in this work are far away from us, i.e. $\sim$ 0.1 kpc, eq. \eqref{TProb_finite_r} can be further simplified as
\es{finite_r0}{
		p_{a  \gamma}^\infty = \lim_{r \to \infty} p_{a  \gamma}(r)
	&\approx \frac{\xi(r_c)^2}{2v_c^2 \text{ sin}^2\tilde{\theta}} \, g_{a \gamma \gamma}^2 B(r_c)^2 L^2.
}
Here we have used the fact that $\lim_{x \to \infty} G(x) = 1$. In particular, when $\theta = \pi/2$, we can get
\es{finite_r1}{
	p_{a  \gamma}^\infty \approx \frac{1}{2v_c^2} \, g_{a \gamma \gamma}^2 B(r_c)^2 L^2,
}
where $L = \sqrt{2\pi v_c H_\text{cor}/m_a}$.

In the left panel of Fig.\ref{figs1}, we show the relationship between $p_{a  \gamma}$ and $r$ for the MWD candidate WD 2010+310 with $\theta = \pi/2$, in which the solid blue line and red dashed line denote the numerical result by solving eq. \eqref{ode}) and the analytical approximation using eq. \eqref{TProb_finite_r}, respectively. we can see that conversion happens in a very narrow region around $r_c$ and converges to the result of eq. \eqref{finite_r1} very quickly. In the right panel of Fig.\ref{figs1}, we demonstrate the relationship between $p_{a  \gamma}$ and $\theta$. It shows that $p_{a  \gamma}$ varies about one order of magnitude with $\theta$ from 0 to $\pi$ \footnote{Note that when $\theta$ is equal to 0 or $\pi$, the direction of axion motion is parallel or antiparallel to the magnetic field and $P_{a\gamma}$ goes to zero; while for a tiny shift of $\theta$, the $P_{a\gamma}$ is restored. In the next section we simply take $\theta = \pi/2$ and ignore these extreme cases.}.
In the next section, we simply set $\theta = \pi/2$ \footnote{Under the simplifying assumption that the magnetic field has a dipolar structure, it is possible to derive the strength and the angle between the observer's line of sight and the magnetic field axis by using the Zeeman shift of a line component, as well as the circular and linear polarisations of a Zeeman split component. More technical details can be found in \cite{2020AdSpR}.}.

It is known that when the electromagnetic wave propagates in the plasma, its amplitude modulates due to the varying plasma frequency $\omega_p(r)$. This effect is not considered in the above calculation. In the case considered here, the amplitude of the outgoing electromagnetic wave $\tilde{A}_y$ decreases with increasing $r$. As shown above, the conversion dominantly takes place in a very narrow region around $r_c$. After conversion, the suppression factor of $\tilde{A}_y$ during propagation can be given as $\sim \tilde{A}_y(r)/\tilde{A}_y(r_c)\sim (m_av_c)^{1/2}/(\omega^2-\omega_p^2(r))^{1/4}$ \cite{1804.03145}. Note that although $\tilde{A}_y$ varies with $\omega_p(r)$ in the plasma, the energy flux of the electromagnetic wave denoted by the magnitude of the Poynting vector remains a constant. Therefore, the suppression of $\tilde{A}_y$ during propagation is not needed to consider for the calculation of energy flux.

\section{The radio flux density from the magnetic white dwarf}
\label{radioflux}

Once converted, the outgoing photons would be absorbed or scattered in the MWDs' coronae, which is characterized by opacity. There are two important processes: the inverse bremsstrahlung process and Compton scattering. The absorption and scattering rate are given as \cite{2010.15836,0801.1527}
\ba\label{eq:brem}
\Gamma_{\rm inv} &\approx& \frac{8\pi n_e Z_N^2 n_N \alpha^3}{3 \omega^3 m_e^2} \left( \frac{2\pi m_e}{T_\text{cor}} \right)^{1/2} \log\left( \frac{2T^2_\text{cor}}{m_\gamma^2} \right) \left(1 - e^{-\omega/T_\text{cor}}\right),
\ea
\ba
	\label{eq:Compton}
	\Gamma_{\rm Com} = \frac{8\pi  \alpha^2}{3  m_e^2}  n_e,
\ea
where $n_N$ is the number density of the charged ions with a charge $e Z_N$\footnote{Since we have assumed that the corona is composed of
	fully ionized hydrogen plasma (see Sec. \ref{structure_WD}), so we have $Z_N = 1$ and $n_N = n_e$ under electric neutral condition, while for the helium plasma, we can get $Z_N = 2$ and $n_N = n_e/2$,  which means $\Gamma_{\rm inv}^\text{He} \simeq 2\Gamma_{\rm inv}^\text{H}$ and does not make much difference to the survival probability $P_s$.}.
Then the survival probability for the converted photons escaping from the MWD can be expressed as
\ba
	P_{s}  \simeq \exp\left[- \int_{r_c}^{\infty}  dr \left(\Gamma_{\rm inv} + \Gamma_{\rm Com}\right) \right].
	\label{eq:Ps}
\ea
For the MWD candidate  WD 2010+310 in Table.\ref{tab:candidates} we get  $P_{s} \sim 0.99$ with the typical parameters $m_a = 10^{-6} \, \mathrm{eV}$ and $g_{a\gamma\gamma} = 10^{-12} \ \mathrm{GeV}^{-1}$. This means that the corona of the MWD is optically thin and the scattering and absorption of the photons can be safely ignored.

It is known that when the photon propagates in the plasma, its amplitude modulates due to the varying plasma frequency, while its energy flux denoted by the Poynting vector remains a constant. It is convenient for us to calculate the radiated power $\mathcal{P}$ in a solid angle $d\Omega$ at $r_c$ as \cite{1804.03145}
\es{dPdOmegarc}{
	{d\mathcal{P} \over d \Omega}  \approx 2 \times p_{a \gamma}^\infty \, \rho_\text{DM}^{r_c} v_c r_c^2 \,,
}
where the factor of two is from the fact that the DM may be converted into photons either on its way in to or out of the resonant layer, and $\rho_{\rm DM}^{r_c}$ is the DM mass density at $r_c$. We denote $\rho_\text{DM}^\infty \sim 0.3 ~\text{GeV}/\text{cm}^3$ as the axion DM density at infinity far away from the MWD candidates, and assume that the axion particles obey the Maxwell-Boltzmann velocity distribution. By using Liouville's theorem, we can map the phase-space distribution from asymptotic infinity to $r_c$.
In the limit $v_0 / v_c \ll 1$, the  $\rho_{\rm DM}^{r_c}$ can be given by \cite{1804.03145}
\es{}{
	\rho_\text{DM}^{r_c} = \rho_\text{DM}^\infty {2 \over \sqrt{\pi}} {v_c \over v_0} + \cdots,
}
where $v_0\sim 200~ \text{km}/\text{s}$ is the DM virial velocity.
The raido flux density at the Earth is given by
\ba
\label{pflux}
S_{a\gamma}=\frac{ d\mathcal{P} }{d\Omega}\frac{1}{ \mathcal{B} d^2},
\ea
where $d$ represents the distance from the MWD to us, $\mathcal{B} = \text{max}\{B_\text{sig}, B_\text{res}\}$ is the optimized bandwidth, $ B_\text{sig}\sim v_0^2 m_a/(2\pi)$ is the signal bandwidth, which is determined by the velocity dispersion in the asymptotic DM distribution, and $B_\text{res}$ is the telescope spectral resolution.
It is worth noting that in general $B_\text{sig}$ is smaller than $B_\text{res}$ (see  Table.~\ref{tab:AeffTsys}).
\begin{table}[!t]
	\centering
	\begin{tabular}{c|cccccc}
		\multicolumn{7}{c}{Parameters and expected radio flux density of the MWDs} \\
		\hline\hline
		& $M_\text{WD} \ [M_\odot]$ & $R_\text{WD} \ [R_\odot]$ & $T_\mathrm{eff} \ [\mathrm{K}]$ & $B \ [\mathrm{MG}]$ & $d_\text{WD} \ [\mathrm{pc}]$ & $S_{a\gamma} \ [\mu\mathrm{Jy}$] \\ \hline
		\mbox{WD 09487-2421} & $0.84$ & $0.0098$ &  $14530$ & $670$ & $36.53$ & $85.33$ \\
		\mbox{WD 2010+310} & $1.14$ & $0.00643$ &  $19750$ & $520$ & $30.77$ & $110.02$ \\
		\mbox{WD 1031+234} & $0.937$ & $0.00872$ & $20000$ & $200$ & $64.09$ & $2.97$ \\
		\mbox{WD 1043-050} & $1.02$ & $0.00787$  & $16250$ & $820$ & $83.33$ & $33.51$ \\
		\mbox{WD 1743-520} & $1.13$ & $0.00681$ & $14500$ & $36$ & $38.93$ & $0.33$ \\
		\hline\hline
	\end{tabular}
	\caption{\label{tab:candidates}MWDs that make good candidates for the detection of the  axion-induced radio flux. The columns correspond to the star's mass in solar mass, radius in solar radius, effective temperature in Kelvin, magnetic field strength in mega-Gauss, distance from the Earth in parsecs, and predicted radio flux density in $\mu\mathrm{Jy}$. Some typical parameters are taken as $m_a = 10^{-6} \, \mathrm{eV}$, $g_{a\gamma\gamma} = 10^{-12} \ \mathrm{GeV}^{-1}$, $n_{e0} = 10^{10}~ \text{cm}^{-3}$, and $T_\text{cor} = 10^6~\text{K}$. With these parameters, $\mathcal{B}$ is derived to be $1.0~ \text{kHz}$. The parameters from observations were obtained by merging the catalogs in Refs.~\cite{1903.05088,Kleinman:2012nt,Ferrario:2015oda,Brown:2018dum}.
	}
	\label{table1}
\end{table}

Using eq. \eqref{finite_r1}, \eqref{dPdOmegarc} $\sim$ \eqref{pflux} we can calculate the radio flux density for the specific MWDs. In Table.\ref{tab:candidates} we list the parameters of five MWD candidates as well as their radio flux densities $S_{a\gamma}$. The following typical parameters are selected for the calculation: $m_a = 10^{-6} \, \mathrm{eV}$, $g_{a\gamma\gamma} = 10^{-12} \ \mathrm{GeV}^{-1}$, $n_{e0} = 10^{10}~ \text{cm}^{-3}$, $T_\text{cor} = 10^6~\text{K}$. With these parameters we can derive that $\mathcal{B} = 1.0~ \text{kHz}$ (see Talbe. \ref{tab:AeffTsys}).

Interestingly enough, from eq. \eqref{resr} we find that $r_c$ is roughly a constant, say $r_c \simeq R_\text{WD}$, by using this simplification we can get an intuitive but approximate result
\begin{align}
   S_{a\gamma}^\text{WD} & \simeq 29.11 ~\mu\text{Jy}\left(\frac{\rho_\text{DM}^\infty}{0.3 ~\text{GeV}/\text{cm}^3}\right)\left(\frac{M_\text{WD}}{M_\odot}\right)^{3/2}\left(\frac{v_0}{200~ \text{km}/\text{s}}\right)^{-1}\left(\frac{R_\text{WD}}{10^4 ~\text{km}}\right)^{-1/2}\left(\frac{T_\text{cor}}{10^6 ~\text{K}}\right)\notag\\
   &\qquad\qquad~~ \times\left(\frac{g_{a\gamma}}{10^{-12}~\text{GeV}^{-1}}\right)^2\left(\frac{B_0}{10^8~\text{G}}\right)^2\left(\frac{m_a}{1~\mu\text{eV}}\right)^{-1}\left(\frac{d}{10~\text{pc}}\right)^{-2}\left(\frac{\mathcal{B}}{1~\text{kHz}}\right)^{-1}.
\end{align}
As a comparison, here we also demonstrate the flux density of NSs \cite{1804.03145}
\begin{align}
	S_{a\gamma}^\text{NS} & \simeq 71.97 ~\mu\text{Jy}\left(\frac{\rho_\text{DM}^\infty}{0.3 ~\text{GeV}/\text{cm}^3}\right)\left(\frac{M_\text{NS}}{M_\odot}\right)^{1/2}\left(\frac{v_0}{200~ \text{km}/\text{s}}\right)^{-1}\left(\frac{R_\text{NS}}{10 ~\text{km}}\right)^{5/2}\left(\frac{P}{1 ~\text{sec}}\right)^{7/6}\notag\\
	&\qquad\qquad~ \times\left(\frac{g_{a\gamma}}{10^{-12}~\text{GeV}^{-1}}\right)^2\left(\frac{B_0}{10^{14}~\text{G}}\right)^{5/6}\left(\frac{m_a}{1~\mu\text{eV}}\right)^{4/3}\left(\frac{d}{100~\text{pc}}\right)^{-2}\left(\frac{\mathcal{B}}{1~\text{kHz}}\right)^{-1},
\end{align}
where $M_\text{NS}$ and $R_\text{NS}$ represent the mass and radius of NSs, and $P$ is the NS spin period. We can see that there are two major differences: (1) the rotation period $P$ goes into the NSs' expression, while for MWD the $T_\text{cor}$ appears; (2) the power index of mass ($M_\text{NS}/M_\text{WD}$), radius ($R_\text{NS}/R_\text{WD}$), $B_0$, and $m_a$ are different. All of these differences are caused by the difference in the plasma distribution functions. Since in this work we use the barometric formula to describe the MWDs' plasma, while for neutron stars the plasma distribution is determined by the GJ model \cite{1804.03145,1969ApJ}. Another interesting fact is that MWDs, though, have a weaker magnetic field than NSs, the advantage of geometric size and spatial distance can compensate it.

All the results should be compared to the minimum detectable flux density $S_{\min}$ of a radio telescope, such as SKA~\cite{ska}. By using the radiometer equation, it can be given by \cite{1804.03145}
\begin{align}
	S_{\min} = \frac{\rm SEFD }{\eta_s \sqrt{n_{\rm pol}  \, {\cal B} ~t_{\rm obs}} }\ ,
	\label{eq:Smin1}
\end{align}
where
\begin{align}
	{\rm SEFD } = \frac{2 k_B }{ A_{\rm eff}/T_{\rm sys} }
	\label{eq:SEFD}
\end{align}
is the system-equivalent flux density, $n_{\rm pol} =2$ is the number of polarization,
$t_{\rm obs}$ is the observation time,
$\eta_s$ is the system efficiency, $k_B$ is the Boltzmann constant, $T_{\rm sys}$ is the antenna system temperature, and
$A_{\rm eff}$ is the antenna effective area of the array. More detailed derivations can be found in Ref. \cite{2016era}. In this work, we take $\eta_s = 0.9$ for SKA~\cite{ska}. The values of the telescope spectral resolution $B_{\rm res}$ for SKA are listed in Table.\ref{tab:AeffTsys}.

\begin{table}[!t]
	\centering
		\begin{tabular}{c|c c c c c c}
			\multicolumn{6}{c}{Parameters and sensitivity of the SKA} \\
			\hline\hline
			Name	&$f$ [MHz]
			& $B_{\rm res}$ [kHz]
			& $ A_{\rm eff}/ T_{\rm sys} $ [${\rm m^2/ \rm K}$] & SEFD [Jy] & $S_\text{min}$ [$\mu$Jy]
			\\
			\hline
			SKA1-Low & (50, 350) & 1.0 & 1000 & 2.76 & 140.0 \\
			SKA1-Mid B1 & (350, 1050)  & 3.9 & 779 & 3.54 & 91.0
			\\
			SKA1-Mid B2 & (950, 1760) &  3.9& 1309 & 2.11 & 54.2\\
			SKA1-Mid B3 & (1650, 3050) &  9.7& 1309 & 2.11 & 34.3\\
			SKA1-Mid B4 & (2800, 5180) &  9.7& 1190 & 2.32 & 37.8\\
			SKA1-Mid B5 & (4600, 13800) &  9.7& 994 & 2.78 & 45.2\\
			\hline\hline
		\end{tabular}
	\caption{The frequency range, telescope spectral resolution  $B_{\rm res}$,
		the ratio between averaged effective area $A_{\rm eff}$ and averaged system temperature $ T_{\rm sys}$, SEFD, and minimum detectable flux density
		in the different frequency bands for SKA1.}
	\label{tab:AeffTsys}
\end{table}

\begin{figure}[!h]
	\centering
	$$\includegraphics[width=0.6\textwidth]{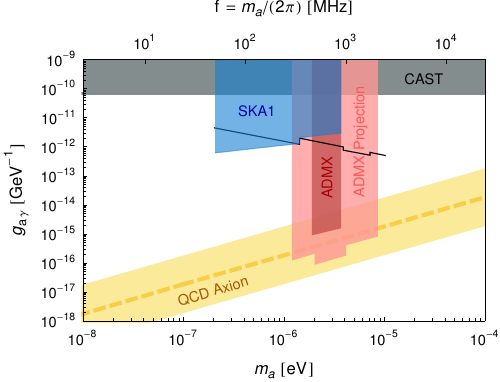}$$
	\caption{The projected sensitivity to $g_{a\gamma}$ as a function of the axion mass $m_a$ for SKA1 telescopes with 100 hours observations of the \mbox{WD 2010+310} is shown in the blue region. The lower mass cutoff is set by the lowest available frequency of SKA, while the upper cutoff is set by requiring the conversion radius to be larger than the MWDs' radius. For comparison, the result of the isolated NS RX J0806.4-4123 is also shown in black solid line. The QCD axion is predicted to lie within the yellow band. The limits set by CAST and ADMX (current and projected) are indicated by the gray and red regions, respectively.}
	\label{fig:1}
\end{figure}

Next we propose to use the SKA phase 1 (SKA1) as a benchmark to search for the radio signals converted from axion DM at MWDs. As shown in Table. \ref{tab:AeffTsys}, it consists of a low-frequency aperture array (SKA1-Low) and a middle frequency aperture array (SKA1-Mid) \cite{ska}. The SKA1-Low covers the $(50, 350)$ MHz frequency band, while the SKA1-Mid actually covers five frequency bands: $(350, 1050)$ MHz, $(950, 1760)$ MHz, $(1650, 3050)$ MHz, $(2800, 5180)$ MHz, and $(4600, 13800)$ MHz. However, considering the constraints of the resonant conversion condition $m_a = \omega_p$ as well as the plasma density $n_{e0} \lesssim 10^{10}~ \text{cm}^{-3}$, there is an upper bound on the frequency of the photon $f_\gamma \lesssim$ 903 MHz. Therefore, we only need to use the first two frequency bands SKA1-Low and SKA1-Mid B1 for MWDs, while for NSs more higher frequency bands are needed (see Fig.\ref{fig:1}). For the sake of completeness, the specific parameters of all low and middle frequency bands of the SKA are listed in Table. \ref{tab:AeffTsys}.

In Fig. \ref{fig:1} we show the sensitivity to $g_{a\gamma}$ for the WD 2010+310, which is the strongest source in Talbe. \ref{tab:candidates}. The blue regions show the physics potential of SKA1 with 100 hours of observation. Note that the lower mass cutoff is set by the lowest available frequency of SKA $\sim$ 50 MHz, while the upper cutoff is set by requiring the conversion radius to be larger than the MWD radius. We find that the axion DM in the mass range of $0.2 \sim 3.7~ \mu$eV can be probed effectively, and the upper limit sensitivity of $g_{a\gamma}$ is $\lesssim 10^{-12} ~\text{GeV}^{-1}$.

For comparison, the result of isolated NS RX J0806.4-4123 is also displayed in black solid line. Comparing with the result in Ref.\cite{1804.03145}, here we use the experimental parameters of SKA in Table. \ref{tab:AeffTsys} to give a more specific result. It shows that the NSs sources can be used to detect a wider range of axion mass (thanks to higher plasma density), while in the low mass region ($m_a \lesssim 1.5 ~\mu\text{eV}$), the WD 2010+310 could give a greater sensitivity.
In addition, the limitations given by the other two experiments, including ADMX and CAST, are also shown in red and gray regions, respectively.

\section{Conclusions}\label{conclusions}
In this work we propose to use MWDs as probes to detect the axion DM through the radio signals. It is known that the MWDs can host very strong magnetic field (e.g. $10^7\sim 10^8$ G). If we adopt the corona parameters that fulfill the X-ray constraints, such as $n_{e0} \sim 10^{10}~ \text{cm}^{-3}$ and $T_\text{cor} \sim 10^6~\text{K}$, the plasma frequency is given by $\omega_p\sim \mu\text{eV}$ (corresponding to the frequency $\sim$ GHz). We find that the resonant conversion may happen when axions pass through
the magnetosphere that is a narrow region around the radius $r_c$, at which the plasma frequency is equal to the axion mass.
Besides, we show that the effects of the inverse bremsstrahlung process and Compton scattering for the outgoing photons are negligible ($P_{s}\sim 0.99$). Therefore, once converted, the radio photon can pass unimpededly through the MWD's corona and be detected by the radio telescope on Earth.

Meanwhile, it is intriguing that for the axion DM with a mass $\sim \mu \text{eV}$, which happens to be in the sensitive region of the terrestrial radio telescopes, such as SKA. In Sec. \ref{radioflux} we use the MWD WD 2010+310 as a target and show the sensitivity to $g_{a\gamma}$ from the future experiment SKA phase 1 with 100 hours of observation. We find that the planned SKA1 can promisingly explore the parameter space of the
axion-photon coupling $g_{a\gamma}$ up to $\sim 10^{-12}~ \text{GeV}^{-1}$ in the axion mass range of $0.2 \sim 3.7~ \mu$eV, which could be more sensitive than NS RX J0806.4-4123 in the low mass region ($m_a \lesssim 1.5 ~\mu\text{eV}$). This result may increase by more than one order of magnitude in the SKA phase 2 (SKA2) \cite{ska,1803.08230}.

Note that all of the MWDs considered are isolated. In fact, one can consider another class of MWDs that occupy regions of high DM density and/or low velocity dispersion, such as the galactic center, dwarf galaxies, and so on. In these regions, the DM density may be enhanced by a large factor. In addition, in dwarf galaxies the velocity dispersion of DM can be low as $v_0 \sim 10~\text{km/s}$. These cases would significantly improve our results and are left for the future work.

\subsubsection*{Acknowledgements}

The authors would like to thank Lilia Ferrario, Anson Hook, Yonatan Kahn, Domitilla de Martino, Alessandro Strumia, Piero Ullio, and Samuel J. Witte for helpful discussions.
The work of JWW is supported by the research grant "the Dark Universe: A Synergic Multi-messenger Approach" number 2017X7X85K under the program PRIN 2017 funded by the Ministero dell'Istruzione, Universit$\grave{a}$ e della Ricerca (MIUR).
The work of XJB, RMY, and PFY is supported by the National Key R\&D Program of China (No. 2016YFA0400200), the National Natural Science Foundation of China (Nos. U1738209 and 11851303).

\footnotesize

\bibliographystyle{utphys}
\bibliography{WDconversion}

\end{document}